\documentclass[aps,pra,reprint,longbibliography,superscriptaddress]{revtex4-2}
\usepackage[T1]{fontenc}
\usepackage[utf8]{inputenc}
\usepackage{graphicx}
\usepackage{amsmath}
\usepackage{xcolor}
\usepackage{txfonts}
\usepackage{times}
\usepackage{hyperref}
\usepackage{orcidlink}
\usepackage{enumitem}   

\definecolor{bggreen}{RGB}{185,230,70}
\definecolor{myblue}{RGB}{0, 40, 140}

\hypersetup{
    unicode=true,          
    pdftitle={Analytic approximations to the strengths of near-threshold optical Feshbach resonances},    
    pdfauthor={Mateusz Borkowski},     
    colorlinks=true,       
    linkcolor=myblue,          
    citecolor=myblue,        
    filecolor=myblue,      
    urlcolor=myblue,           
    bookmarks=false
}

\begin{document}
	\title{Analytic approximations to the strengths of near-threshold optical Feshbach resonances}
	
	\author{Mateusz Borkowski\orcidlink{0000-0003-0236-8100}}
    \affiliation{Department of Physics, Columbia University, 538 West 120th Street, New York, NY 10027-5255, United States of America}
    \affiliation{Van der Waals-Zeeman Institute, Institute of Physics, University of Amsterdam, Science Park 904, 1098 XH Amsterdam, The Netherlands}
    \affiliation{Institute of Physics, Faculty of Physics, Astronomy and Informatics, Nicolaus Copernicus University, Grudziadzka 5, 87-100 Torun, Poland}
	
	\date{\today}
	
\begin{abstract}
Optical Feshbach resonances (OFRs) allow one to control cold atomic scattering, produce ultracold molecules and study atomic interactions via photoassociation spectroscopy. In the limit of ultracold s-wave collisions the strength of an optical Feshbach resonance can be expressed via an energy-independent parameter called the optical length. Here we give fully analytic approximate expressions for its magnitude applicable to near-threshold bound states of an excited molecular state dominated by a single resonant-dipole or van der Waals interaction. We express these magnitudes in terms of intuitive quantities, such as the laser intensity, excited state binding energy, the s-wave scattering length and the Condon point. Additionally, we extend the utility of the optical length to associative STIRAP in 3D optical lattices by showing that the free-bound Rabi frequency induced by a laser coupling a pair of atoms in an optical lattice site can be approximately related to the trap frequency and the optical length.
\end{abstract}
\maketitle
 
\section{Introduction}

Feshbach resonances~\cite{Chin2010} emerge in cold atomic collisions when the entrance scattering channel is coupled to a discrete molecular state by e.g. hyperfine mixing. Optical Feshbach resonances (OFRs)~\cite{Jones2006} are created artificially by a laser tuned nearby an electronically excited molecular bound state. The unavoidable losses due to spontaneous decay from the excited state are the foundation of photoassociation spectroscopy: an~essential tool for studying weakly bound states in homonuclear \cite{Thorsheim1987,Lett1993, Miller1993, Ratliff1994, Zinner2000, Zelevinsky2006, Borkowski2011, Takasu2012, Borkowski2014, Kim2016, Reschovsky2018} and heteronuclear molecules composed of the colliding species \cite{Munchow2011, Ciamei2018, Roy2016, Guttridge2018}. For several systems, most notably $^{88}$Sr~\cite{Reinaudi2012},  the decay from the excited bound state may efficiently produce ultracold ground state molecules~\cite{Sage2005, Deiglmayr2008, Reinaudi2012, Quemener2012, Bruzewicz2014, Borkowski2017b}. 
Finally, OFRs can control the scattering length \cite{Fedichev1996, Bohn1997, Fatemi2000, Theis2004, Thalhammer2005, Enomoto2008, Yamazaki2010a, Blatt2011, Yan2013a} with high spatial and temporal resolution \cite{Yamazaki2010a}. 

\begin{figure}[b]
    \begin{center}
        \includegraphics
        {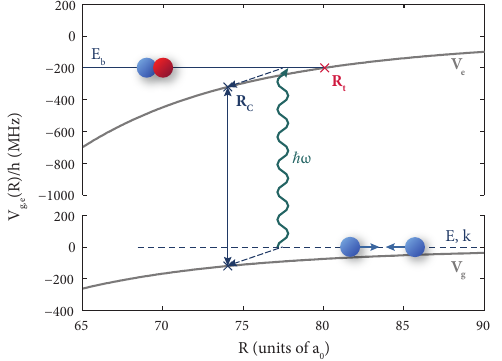}
    \end{center}
    \caption{In an optical Feshbach resonance the initial atomic scattering state, characterized by a collision energy $E$ and wavevector $k$ are coupled by the incident photon $\hbar \omega$ to an excited molecular bound state at energy $E_b$. Here $V_g$ and $V_e$ are respective ground- and excited state potential, $R_t$ is the classical outer turning point of the excited bound state, and $R_C$ is the outermost Condon point: the internuclear distance where the photon energy matches exactly the energy difference between the ground and excited state potentials.}
    \label{fig:schematic}
\end{figure}

The strength of an $s$-wave OFR can be expressed by an optical length $l_{\rm opt}$~\cite{Bohn1997, Ciurylo2005, Jones2006, Ciurylo2006} related to a free-bound Franck-Condon factor between the ground state scattering and excited bound state wavefunctions. These can be evaluated numerically by solving the appropriate radial Schr\"odinger equations. Alternatively the Franck-Condon factors can be approximated using the well known reflection approximation~originally devised by Jabłoński in the context of his quantum theory of spectral line broadening~\cite{Jablonski1945} and later appropriated for cold collisions by Julienne~\cite{Julienne1996}. The major virtue of the reflection approximation is that it relates the strength of a photoassociation line to the squared ground state wavefunction at the outermost Condon point (Fig.~\ref{fig:schematic}) -- a point where the incident photon's energy matches the energy difference between ground and excited state potential -- and thus can e.g. explain gaps in photoassociation spectra as due to nodes in the ground state wavefunction. The reflection approximation works well when the excited state potential has a much longer range than the ground state potential, for instance when the excited state is dominated by a strong resonant-dipole (r-d) interaction. An improved version of the reflection formula, the ``stationary phase approximation''~\cite{Ciurylo2006} has an extended range of validity that also covers van der Waals (vdw) excited states so long as the excited state vdW interaction is much stronger than the ground state's. In practice, however, to use either formula one still must compute the ground state wavefunction numerically~(e.g. \cite{Johnson1977, Colbert1992, Tiesinga1998}), estimate the excited state vibrational splittings, and, for the stationary phase formula, numerically integrate a WKB phase shift, often lacking knowledge of the relevant interatomic potentials. This may require postulating a model potential without prior experimental input. Then, calculating e.g. the dependence of the photoassociation spectra on the scattering length would require further manual tuning of that potential.

In this paper we evaluate the reflection~\cite{Julienne1996, Boisseau2000} and stationary phase formulas~\cite{Ciurylo2006} (where appropriate) using the WKB approximation to arrive at fully analytic formulas for the strengths of optical Feshbach resonances that require no numerical solutions to the radial Schr\"odinger equation. To do so, we will consider the excited-state interaction potential $V_e(R)$ as dominated by a single $-C^e_n R^{-n}$ term, where $R$ is the internuclear distance. In particular, we will consider two useful limiting cases:
\begin{enumerate}[label=(\roman*)]
\item an excited state with a strong resonant-dipole (r-d) interaction $V_e\sim -C_3^e R^{-3}$, which describes homonuclear collisions in a laser field tuned near either an allowed or sufficiently strong intercombination transition, and
\item a van der Waals (vdw) excited state $V_e \sim -C_6^e R^{-6}$, which describes heteronuclear systems.
\end{enumerate}
We will express $l_{\rm opt}$ using intuitive physical quantities: the $s$-wave scattering length~$a$, leading interaction terms ($C_3^e$ or $C_6^e$), the excited state binding energy $E_b$, the Condon point $R_C$, and classical outer turning point $R_t$. In particular, the explicit connection to the scattering length could help experimentalists choose the best isotopologue of a system via mass-scaling~\cite{Gribakin1993,Verhaar2009,Borkowski2013}. We will test our formulas on real-world examples: intercombination line OFRs in Yb$_2$ (a system whose excited state interactions are dominated by a resonant-dipole $R^{-3}$ term)~\cite{Tojo2006, Enomoto2008, Borkowski2009, Kim2016}, and OFRs in the Rb+Sr system near the Rb D1 line~\cite{Devolder2018} (which has a van der Waals $R^{-6}$ tail in the excited state interaction potential). Despite approximating real atomic interactions with just one dominant $C_n R^{-n}$ term, we will find that the resultant fully analytic formulas can still match numerical computations semi-quantitatively.

Finally, given the interest in the production of molecules from atomic pairs in 3D optical lattice sites~\cite{Stellmer2012, Ciamei2017}, we will show how the ``free-bound'' Rabi frequency $\Omega_{\rm FB}$ for the transition of an unbound atomic pair state in an optical lattice site to the excited molecular bound state can be calculated from the optical length $l_{\rm opt}$ and then test it using $^{84}$Sr$_2$ experimental data~\cite{Ciamei2017, Reschovsky2018}. Our formula, based on ideas from quantum defect theory~\cite{Mies1984}, corroborates the empirical observation~\cite{Ciamei2017} that the free-bound Rabi frequency scales approximately with the optical lattice frequency $\omega_{\rm trap}$ as $\Omega_{\rm FB}\propto \omega_{\rm trap}^{3/4}$.

\section{Derivation of the analytic formulas}

Here we analytically evaluate optical lengths within the reflection approximation~\cite{Julienne1996, Bohn1999} and its more accurate version, the stationary phase approximation~\cite{Ciurylo2006} by using the WKB approximation. In Sec.~\ref{sec:optlen} we will define the optical length in terms of a free-bound Franck-Condon factor and in Sec.~\ref{sec:approx} we will quote the reflection and stationary phase approximations for that FCF. In the following sections we will look at the individual terms that comprise these approximations and evaluate them one-by-one. For the ratio of vibrational spacings and potential slope (Sec.\ref{ssec:spacing}) we shall employ the Le Roy-Bernstein formula~\cite{Leroy1970} allowing us to write them in terms the dominant excited state interaction and the excited bound state position. Then, in Sec.~\ref{sec:wfs} we will introduce the three models of the ground state wavefunction in the zero-energy limit: \emph{asymptotic}, \emph{long-range} and \emph{short-range} that allow us to write the wavefunction in terms of the ground-state $s$-wave scattering length and the ground state van der Waals coefficients $C_6^g$ and optionally $C_8^g$. In Sec.~\ref{sec:phasecorr} we will tackle the extra phase correction term that differentiates between the reflection~\cite{Julienne1996} and stationary phase~\cite{Ciurylo2006} approximations by using the WKB approximation on locally linearized excited state potential. In Sec.~\ref{sec:final} we list the final analytic formulas: Eqs.~(\ref{eq:rd_asym})--(\ref{eq:vdw_short}).

\subsection{The optical length~\label{sec:optlen}}

Consider an ultracold collision of two atoms, in $s$-wave, at a kinetic energy $E$, and described by an energy-normalized scattering wavefunction $f_g$.
The strength of an OFR, or its ability to induce observable photoassociative losses or a useful change to the scattering length, is commonly expressed via an ``optical length'', defined as~\cite{Bohn1997, Ciurylo2005, Jones2006, Ciurylo2006},
\begin{equation}
    l_{\rm opt} = \Gamma_{\rm stim}/2k\gamma_m.
\end{equation} 
Here $\gamma_m$ is the decay rate of the excited bound state, $\hbar \Gamma_{\rm stim}=2\pi \rvert \langle f_g \rvert V^{\rm opt} \rvert \psi_b \rangle \rvert^2$ is a stimulated rate~\cite{Bohn1999} induced by coupling an excited bound state $\psi_b$ to the ground scattering wavefunction $f_g$, via a matrix element $V_{\rm opt}$ that describes the interaction of the colliding atoms with light. For a laser detuned by $\delta$, the change to scattering length $\Delta a$ and the photoassociative inelastic collision rate $K_{\rm in}~$\cite{Bohn1997,Jones2006, Blatt2011, Yan2013, Nicholson2015, Kim2016, Reschovsky2018} are
\begin{eqnarray}
    \Delta a & = & \frac{l_{\rm opt}\gamma_m \delta}{\delta^2+(\eta \gamma_m)^2/4},\,{\rm and} 
    \\ K_{\rm in} & = & g\frac{2\pi\hbar}{\mu} \frac{l_{\rm opt} \eta \gamma_m^2}{\delta^2+\gamma_m^2(\eta +2k l_{\rm opt})^2/4},
\end{eqnarray}
where $\gamma_m$ is the natural linewidth of the excited bound state, $\eta \geq 1$ is a broadening factor that accounts for other loss processes, $\mu$ is the reduced mass, $g$ is a symmetry factor (2~for a thermal gas of identical bosons, 1 otherwise) and the wavenumber $k=\sqrt{2\mu E}/\hbar$ at collision energy $E$. 

In this paper we consider optical Feshbach resonances close to near-threshold excited molecular bound states where the molecular dipole transition moment and the molecular state decay rate $\gamma_m$ can be expressed in terms of the properties of the constituent atoms. In such case it is well known~\cite{Napolitano1997, Bohn1999, Machholm2001, Ciurylo2004, Ciurylo2005, Borkowski2009, Nicholson2015} that the optical length becomes
\begin{equation}
    l_{\rm opt} =  \frac{3\lambda_a^3}{16 \pi c} I f_{\rm rot} \frac{\rvert \langle f_g \rvert \psi_b \rangle \rvert^2}{k}, \label{eq:lopt}
\end{equation}
where $I$ is the laser intensity (hereafter assumed equal to 1~$\rm W/cm^2$), $\lambda_a$ is the atomic transition wavelength, and $f_{\rm rot}$ is a rotational H\"onl-London factor stemming from translating electric field operator from space-fixed to body-fixed coordinates~\cite{Napolitano1997, Brown2003}. Within the Wigner threshold regime the amplitude of the ground-state scattering wavefunction $f_g$ is proportional to $\sqrt{k}$ making $l_{\rm opt}\sim f_g^2/k$ practically constant with respect to collision energy. Thus $l_{\rm opt}$ may be evaluated in the limit of zero energy and used for all collision energies in a sufficiently cold gas~\cite{Nicholson2015}. 

\subsection{The reflection and stationary phase approximations\label{sec:approx}}

The starting point for the present paper is the stationary phase approximation formula~\cite{Julienne1996, Bohn1999, Ciurylo2006} for the Franck-Condon factor which, in this paper, we will further evaluate using WKB approximations. The full derivation of the stationary phase formula can be found in Ref.~\cite{Ciurylo2006}; here we will only briefly outline the derivation. 

The regular wavefunctions for the ground and excited state potentials can be written in the Milne form:
\begin{eqnarray}
\psi_b(R) & = & 
    \left(\frac{\partial E_b}{\partial v}\right)^{1/2}
    \left(\frac{2 \mu}{\pi \hbar^2} \right)^{1/2}
    \alpha_e(R) \sin[\beta(R)], \\
f_g(R) & = & 
    \left(\frac{2 \mu}{\pi \hbar^2} \right)^{1/2}
    \alpha_g(R) \sin[\phi(R)], 
\end{eqnarray}
where $\alpha_g(R)$ and $\alpha_e(R)$ are the respective local amplitudes of the ground and excited state wavefunctions;  $\phi(R)$ and $\beta(R)$ are their respective phases. Thus, the desired Franck-Condon factor, 
\begin{equation}
    \rvert \langle f_g \rvert \psi_b \rangle \rvert^2 = 
        \frac{\partial E_b}{\partial v}
        \left( \frac{2\mu}{\pi \hbar^2} \right)^2
        \left|I_{eg}\right|^2,
\end{equation}
can be written in terms of the following integral:
\begin{equation}
    I_{eg} = \int_0^\infty dR \alpha_e(R) \alpha_g(R) \sin[\beta(R)] \sin[\phi(R)]\,.
\end{equation}
The oscillating terms can be rewritten as
$\sin[\beta(R)] \sin[\phi(R)] = (\cos[\beta(R)-\phi(R)]-\cos[\beta(R)+\phi(R)])/2$. The fast oscillating second term contributes little to the integral and can be omitted.

The crux of the stationary phase approximation lies in the observation that the main contribution to this integral stems from the stationary phase point where $\partial\beta(R)/\partial R - \partial\phi(R)/\partial R \approx 0$. Within the WKB approximation the respective derivatives are $\beta(R)/\partial R \approx \sqrt{2\mu[E_b - V_e(R)]} = k_e(R)$ and $\phi(R)/\partial R \approx \sqrt{-2\mu[V_g(R)]} = k_g(R)$, and the stationary point becomes the Condon point (Fig.~\ref{fig:schematic}), defined as the outermost point $R_C$ where $V_e(R_C) - E_b = V_g(R_C)$. Additionally, while the entirety of the ground state wavefunction can not be described by the WKB approximation (see Sec.~\ref{sec:wfs}), the phase and amplitude of the excited state wavefunction for $R<R_t$ can be well approximated by $\alpha_e(R) = (2\mu[E_b - V_e(R)])^{-1/4}$ and $\beta(R) = -\pi/4-\Delta\beta(R, R_t)$ with the phase integral $\Delta\beta(R, R_t) = \int_R^{R_t} dR' k_e(R')$. $R_t$ is the classical outer turning point.

In the final step of the derivation, the phase difference is expanded to second order around the Condon point $R_C$:
\begin{equation}
    \phi(R) - \beta(R) \approx b_0 + b_1(R-R_C) + \frac{b_2}{2}(R-R_C)^2
\end{equation}
where $b_0 = \phi(R_C) + \Delta\beta(R_C, R_t) + \pi/4$, $b_1 \approx 0$ and $b_2 \approx \mu D_C/\hbar^2 k_e(R_C)$~\cite{Ciurylo2006}. Here the slope difference at the Condon point $D_C = V_e'(R_C)-V_g'(R_C)$. Finally, the integral can be evaluated assuming that the amplitudes $\alpha_{g,e}$ are slowly varying around the Condon point~\cite{Bohn1999, Ciurylo2006}:
\begin{eqnarray}
    I_{eg} & \approx & \frac{1}{2}\int_0^{\infty} dR \alpha_e(R)\alpha_g(R)\cos\left[ 
        b_0 + \frac{b_2}{2}(R-R_C)^2 \right] \nonumber \\
    & \approx & \frac{1}{2}\alpha_e(R_C) \alpha_g(R_C) \sqrt{\frac{2\pi}{b_2}} \cos\left[ b_0 + \pi/4\right] \nonumber \\
    & \approx & - \sqrt{\frac{\pi \hbar^2}{2\mu D_C}}\alpha_g(R_C)\sin\left[\phi(R_C) + \Delta\beta(R_C, R_t) \right].
\end{eqnarray}

The final approximate Franck-Condon factor can be written as a product of four terms:~\cite{Ciurylo2006}:
\begin{eqnarray}
    \rvert \langle f_g \rvert \psi_b \rangle \rvert^2 & \approx &
        \frac{\partial E_b}{\partial \nu} \times
        \frac{1}{D_C} \times
        \left|f_g(R_C)\right|^2 \nonumber \\
        && \times \, \frac{\sin^2[\phi(R_C)+\Delta\beta(R_C, R_t)]}{\sin^2[\phi(R_C)]}
        .\label{eq:spa}
\end{eqnarray}
In this work we will aim to analytically approximate these terms. The first term, $\partial E_b/\partial \nu$, is the local vibrational spacing in the excited state. The second depends on the difference $D_C$ between the excited- and ground-state potential slopes taken at the Condon point $R_C$: $D_C = V_e'(R_C)-V_g'(R_C)$. We will treat the two together in section~\ref{ssec:spacing}. The third term is the squared ground state wavefunction at the Condon point. The last is a phase correction term~\cite{Ciurylo2006} that will improve our model for the more deeply bound states where the difference between $R_C$ and the classical outer turning point $R_t$ can be substantial. For very weakly bound states of a resonant-dipole system the Condon point $R_C$ is very close the outer turning point $R_t$ and Eq.~(\ref{eq:spa}) reduces to the well known reflection approximation~\cite{Jablonski1945, Julienne1996, Bohn1999, Boisseau2000}:

\begin{equation}
    \rvert \langle f_g \rvert \psi_b \rangle \rvert^2 \approx 
        \frac{\partial E_b}{\partial \nu} \times
        \frac{1}{D_C} \times
        \left|f_g(R_C)\right|^2 .\label{eq:ra}
\end{equation}
The stationary phase approximation works best when the two molecular potentials $V_e$ and $V_g$ are very different: while in the r-d case this is usually true unless the r-d interaction is very weak, for vdw systems this implies an excited state interaction coefficient $C_6^e$ significantly larger than the ground state $C_6^g$~\cite{Ciurylo2006}.

\subsection{Vibrational spacing and potential slopes \label{ssec:spacing}}

In our quest towards simple expressions we will benefit from two main observations. Firstly, the Leroy-Bernstein theory~\cite{Leroy1970, LeRoy1973} provides a formula for the vibrational spacing in a $-C_nR^{-n}$ potential:
\begin{equation}
    \frac{\partial E_b}{\partial \nu} = 
        \hbar \sqrt{\frac{2\pi}{\mu}} \frac{\Gamma(1+1/n)}{\Gamma(1/2+1/n)} 
        \frac{n}{C_n^{1/n}} (-E_b)^{(n+2)/2n}\,,
\end{equation}
where $\Gamma(x)$ is the Euler gamma function and $E_b$ is the resonance position. When the OFR laser is on resonance and the collision energy $E\to 0$, the difference in potentials at the Condon point matches the bound state energy, $V_e(R_C)-V_g(R_C) = E_b$. We can write the potential difference in terms of multipole expansions, $V_e-V_g = -C_3^e R^{-3}-\Delta C_6R^{-6}-\ldots$, where $\Delta C_6 = C_6^e-C_6^g$. To the lowest order, the appropriate Condon points for the r-d and vdw systems are $R_C^{\rm r-d}\approx (C_3^e)^{1/3}(-E_b)^{-1/3}$ and $R_C^{\rm vdW}\approx \Delta C_6^{1/6}(-E_b)^{-1/6}$. Similarly, we may approximate the difference in potential slopes with $D_C^{\rm r-d} \approx -3C_3R_C^{-4}$ and $D_C^{\rm vdW} \approx -6 \Delta C_6 R_C^{-7}$. With these choices the first two terms in Eq.~(\ref{eq:spa}) simplify to
\begin{subequations}
    \begin{eqnarray}
        &&\left[\frac{\partial E_b}{\partial \nu} \frac{1}{D_C} \right]^{\rm r-d} \approx  \hbar \sqrt{\frac{2\pi}{\mu}}
            \frac{\Gamma(4/3)}{\Gamma(5/6)}\frac{1}{\sqrt{-E_b}},\, {\rm and} \label{eq:spacing_rd}
            \\
        &&\left[\frac{\partial E_b}{\partial \nu} \frac{1}{D_C}\right]^{\rm vdW} \approx  \hbar \sqrt{\frac{2\pi}{\mu}}
            \frac{\Gamma(7/6)}{\Gamma(2/3)} \left(\frac{\Delta C_6}{C_6^e}\right)^{1/6} \frac{1}{\sqrt{-E_b}}.\label{eq:spacing_vdw}
    \end{eqnarray}
\end{subequations}

\subsection{Zero-energy ground state wavefunctions \label{sec:wfs}}

\begin{subequations}
The second simplification stems from the Wigner threshold law \cite{Wigner1948, Jones1999, Jones2006, Borkowski2009, Nicholson2015}: for sufficiently low collision energies, the ratio ${\rvert \langle f_g \rvert \psi_b \rangle \rvert^2}/{k}$ is effectively constant allowing us to evaluate $f_g$ at zero energy, where simple analytic models are available. However, due to the breakdown of the WKB approximation near $R_{\rm vdW}=(2\mu C_6^g/\hbar^2)^{1/4}/2$~\cite{Boisseau2000, Jones2006} we will be forced to use separate wavefunction models for the ``long-range'' ($R \gtrapprox R_{\rm vdW}$) and ``short-range'' ($R \lessapprox R_{\rm vdW}$) internuclear separations. 

The asymptotic ($R \to \infty$) form of the $s$-wave scattering wavefunction [shown in Fig.~\ref{fig:gndwf}(a)] is
\begin{equation}    
    f_g^{\rm asym}(R) \sim \sqrt{\frac{2\mu}{\pi \hbar^2 k}}\sin(kR+\eta),\label{eq:fg_asym}
\end{equation}
where the phase shift $\eta$ due to the short range potential defines the scattering length via $a = \lim_{k \to 0} -\eta/k$. At internuclear distances closer to $R_{\rm vdW}$ the long range van der Waals interaction causes the wavefunction to deviate from its asymptotic form. Reference~\cite{Julienne1996} gives the following approximate formula:
\begin{equation}
    f_g^{\rm long}(R) \approx \sqrt{\frac{2\mu}{\pi \hbar^2 k}}\sin\left[k\left(R-a-\frac{8}{15}\frac{R_{\rm vdW}^4}{R^3}\right)\right]\left(1-\frac{4}{5}\frac{R_{
    \rm vdW}^4}{R^4}\right).\label{eq:fg_long}
\end{equation}
which closely matches the numerical wavefunction for $R$ ranging from $+\infty$ down to about $R_{\rm vdW}$  [Fig.~\ref{fig:gndwf}(a)].

\begin{figure}
    \begin{center}
        \includegraphics[width=0.48\textwidth]{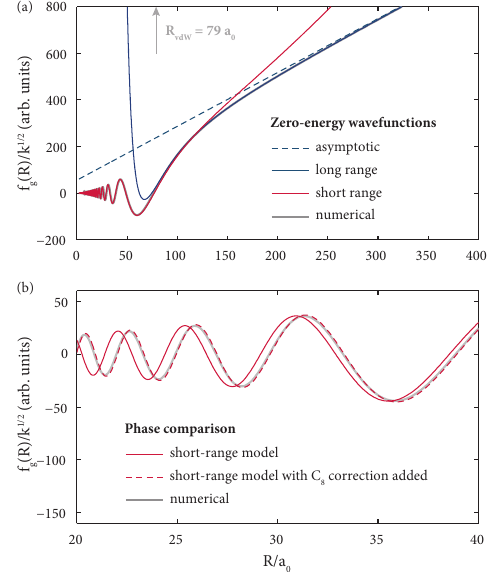}
    \end{center}
    \caption{Comparison of model zero-energy scattering wavefunctions on the example of $^{176}$Yb$_2$ ($a=-24\,a_0$~\cite{Kitagawa2008, Borkowski2017a}). Panel (a) shows the numerical wavefunction (thick grey line) being approximated by respectively the asymptotic wavefunction [Eq.~(\ref{eq:fg_asym}), dashed blue line], its improved version, the long-range model of Ref.~\cite{Julienne1996}, Eq.~(\ref{eq:fg_long}) (solid blue line), and finally the WKB short-range wavefunction~[Eq.~(\ref{eq:fg_short}), red solid line]. In panel (b) we compare the short range WKB wavefuction, Eq.~(\ref{eq:fg_short}), where the wavefunction phase $\phi(R)$, is calculated without (red solid line) and with (red dashed line) the $\phi_8(R)$ phase correction, Eq.~(\ref{eq:phi8}). A quick glance reveals that including the extra phase correction improves the agreement with the numerically obtained wavefunction.}
    \label{fig:gndwf}
\end{figure}

For ``short-range'' interatomic distances, ie. $R \lessapprox R_{\rm vdW}$, we will use a WKB wavefunction
\begin{equation}
    f_g^{\rm short}(R) \approx \sqrt{\frac{2\mu}{\pi \hbar^2}} A(R,E)C^{-1}(E)\sin\left[\phi(R, E)\right],\label{eq:fg_short}
\end{equation}
\end{subequations}
where $A(R,E) = 1/\sqrt{k_{\rm local}(R)}$ and $\phi(R,E)$ are the typical WKB amplitude and phase. Since we assumed that the ground state potential $V_g(R) \sim -C_6^gR^{-6}$, the local wavenumber $k_{\rm local} = [k^2-2\mu V(R)/\hbar^2]^{1/2}$ entering the expression for the WKB amplitude can be replaced with $k_{\rm local} \approx [2\mu C_6^{\rm g}/\hbar^2]^{1/2} R^{-3} = 4 R_{\rm vdW}^2 R^{-3}$.
The additional term $C^{-1}(E) = [k\bar a (1+(a/\bar a-1)^2)]^{1/2}$ is a correction to the amplitude for near-threshold scattering wavefunctions~\cite{Julienne1989, Mies2000, Jones2006}. The quantity $\bar a = 2^{-1/2}[\Gamma(3/4)/\Gamma(5/4)]R_{\rm vdW} = 0.956\ldots\times R_{\rm vdW}$ is the ``mean scattering length'' that enters the semiclassical formula for $a$ in a vdw potential~\cite{Gribakin1993}. 

The zero-energy WKB phase $\phi(R)$ can be related to the $s$-wave scattering length at its large-$R$ limit, $\phi_\infty = \int_{R_0}^\infty k_{\rm local}(R')dR'$, that enters the well-known semiclassical formula, $a=\bar a\left[1-\tan(\phi_\infty-3\pi/8)\right]$~\cite{Gribakin1993}. Thus we can very well start at infinite nuclear separation with the asymptotic value of $\phi_{\infty}$ obtained by inverting the semiclassical formula and accumulate the WKB phase \emph{inwards}. Since the Condon points $R_C$ are well past the LeRoy radius~\cite{LeRoy1973}, we can expand the ground state potential as $V_g\approx-C_6^gR^{-6}-C_8^g R^{-8}$ and express the phase $\phi(R)$ as
\begin{equation}
    \phi(R) = \phi_\infty-\phi_6(R)-\phi_8(R)\,, \label{eq:phi}
\end{equation}
where the individual phase terms are
\begin{subequations}
\begin{eqnarray}
    \phi_\infty &=& 3\pi/8+\arctan\left(1-a/\bar a\right), \\
    \phi_6(R) &=& \int_R^\infty\left(\frac{2\mu C_6^g}{\hbar^2}R^{-6}\right)^{1/2} dR = 
        2\left(\frac{R_{\rm vdW}}{R}\right)^2,\\
    \phi_8(R) &=& \frac{1}{8}\frac{\sqrt{2\mu}}{\hbar} \frac{C_8^g}{\sqrt{C_6^g}}R^{-4}. \label{eq:phi8}
\end{eqnarray}
\end{subequations}
The term $\phi_6$ is the exact WKB phase due to just the $R^{-6}$ tail. The next van der Waals term, $-C_8^gR^{-8}$ can be accounted for perturbatively. A first order WKB phase correction due to a small additional potential $\delta V$ on top of a potential $V$ may be expressed as ~\cite{Lutz2016}
\begin{equation}
\delta \phi \approx \frac{1}{2}\int_{R_C}^{\infty}\frac{\sqrt{2\mu}}{\hbar}\frac{\delta V}{\sqrt{|V(R)|}} dR.
\end{equation} The expression for $\phi_8$ is obtained by assuming the $\delta V = -C_8R^{-8}$ contribution is much smaller than the $V = -C_6R^{-6}$ term. While $\phi_6$ alone may be sufficient for many applications, taking the $C_8$ term into account significantly improves the model wavefunction at shorter internuclear separations, as shown in Fig.~\ref{fig:gndwf}(b). 

\begin{widetext}
Finally in the zero-energy ($k\to 0$) limit we have
\begin{subequations}
\begin{eqnarray}
    \left[\frac{f_g^2(R_C)}{k}\right]^{\rm asym} \,& \approx & {\frac{2\mu}{\pi \hbar^2}} (R_C-a)^2, \label{eq:wf_asym}\\
    \left[\frac{f_g^2(R_C)}{k}\right]^{\rm long} \,& \approx & {\frac{2\mu}{\pi \hbar^2}} \left(R_C-a-\frac{8}{15}\frac{R_{\rm vdW}^4}{R_C^3}\right)^2\left(1-\frac{4}{5}\frac{R_{\rm vdW}^4}{R_C^4}\right)^2 \label{eq:wf_long}\\
    \left[\frac{f_g^2(R_C)}{k}\right]^{\rm short} & \approx & {\frac{2\mu}{\pi \hbar^2}} \bar a \left[1+ (a/{\bar a}-1)^2\right]\,  
     \times \frac{R_C^3}{4R_{\rm vdW}^2} \sin^2\left[\phi(R_C)\right] \,. \label{eq:wf_short}
\end{eqnarray}
\end{subequations}
\end{widetext}
We note that while the simple asymptotic expression only reproduces the outermost node of the scattering wavefunction (and only for a large positive scattering length), it will turn out useful for transitions to weakly bound states supported by a strong \mbox{r-d} interaction whose Condon points are usually well past $R_{\rm vdW}$. The short range model is a rapidly oscillating function due to the $\sin^2[\phi(R_C)]$ term and is appropriate for vdw systems and more deeply bound states in the r-d case. The typical values of $R_{\rm vdW}$ range from about 30~$a_0$ to about 100~$a_0$~\cite{Jones2006}.

\subsection{WKB phase correction \label{sec:phasecorr}}
The last term in Eq.~(\ref{eq:spa}) adds an excited-state WKB phase correction $\Delta \beta(R_C; R_t)=\int_{R_C}^{R_t}[2\mu(E_b-V_e)/\hbar^2]^{1/2}dR$~\cite{Ciurylo2006} that improves upon the reflection approximation for deeply bound states. For this reason, we will only use it in conjunction with the short-range wavefunction model [Eq.~(\ref{eq:fg_short})]. We can obtain an approximate analytic solution to this integral by simply linearizing $V_e$ around the classical outer turning point $R_t$. Since by definition at the turning point $V_e(R_t) = E_b$ we can approximate $V_e(R)$ as $V_e(R) \approx E_b + V_e'(R_t)(R-R_t)$. For the resonant-dipole case
\begin{subequations}
\begin{eqnarray}
    \left[\Delta \beta\right]^{\rm r-d} & \approx & \frac{\sqrt{2\mu C_3^e}}{\hbar}\frac{2\sqrt{3}}{3}R_t^{-2}(R_t-R_C)^{3/2}, \label{eq:phase_corr_rd} 
\end{eqnarray}
whereas for a van der Waals system,
\begin{eqnarray}
    \left[\Delta \beta\right]^{\rm vdW} &\approx &\frac{\sqrt{2\mu C_6^e}}{\hbar}\frac{2\sqrt{6}}{3}R_t^{-7/2}(R_t-R_C)^{3/2}.  \label{eq:phase_corr_vdw}
\end{eqnarray}
\end{subequations}

The distance between the turning and Condon points itself, $R_t - R_C$, can be estimated as follows. The turning point is defined as a point where the excited state potential matches the bound state energy: $V_e(R_t) = E_b$. The position of the Condon point additionally depends on the ground state potential: $V_e(R_C) - V_g(R_C) = E_b$. In a vdw system, if one ignores any potential terms other than $C_6$, then $R_t = (-C_6^e/E_b)^{1/6}$ and $R_C = (-\Delta C_6/ E_b)^{1/6}$. For a r-d system with a dominant $-C_3^e R^{-3}$ excited state interaction we can start from the definitions of $R_C$ and $R_t$ to obtain, after some algebra, 
\[
R_t = R_C\left(1 - \frac{C_6^g R_C^{-3}}{C_3^e}\right)^{-1/3}
\approx R_C\left(1 + \frac{C_6^g R_C^{-3}}{3 C_3^e}\right),
\]
hence $R_t-R_C \approx C_6^gR_C^{-3}/3C_3^e$. Finally we can, without losing much accuracy, replace the $R_C$ on the right hand side with $R_t$ so that $R_t-R_C \approx C_6^gR_t^{-3}/3C_3^e$.

\subsection{Final formulas \label{sec:final}}

Combining the formulas for resonant-dipole vibrational spacing [Eq.~(\ref{eq:spacing_rd})] with, respectively, the asymptotic [Eq.~(\ref{eq:wf_asym})] and long-range wavefunction models [Eq.~(\ref{eq:wf_long})] produces the following approximate expressions:
\begin{widetext}
\begin{subequations}
\begin{eqnarray}
    l_{\rm opt}^{\rm r-d, asym} & = & \frac{3\lambda_a^3}{16\pi c} I f_{\rm rot} \frac{2\sqrt{2\mu}}{\hbar \sqrt{\pi}} \frac{\Gamma(4/3)}{\Gamma(5/6)} \frac{1}{\sqrt{-E_b}} (R_C-a)^2, \label{eq:rd_asym}\\
    l_{\rm opt}^{\rm r-d, long} & = & \frac{3\lambda_a^3}{16\pi c} I f_{\rm rot} \frac{2\sqrt{2\mu}}{\hbar \sqrt{\pi}} \frac{\Gamma(4/3)}{\Gamma(5/6)} \frac{1}{\sqrt{-E_b}} \left(R_C-a-\frac{8}{15}\frac{R_{\rm vdW}^4}{R_C^3}\right)^2\left(1-\frac{4}{5}\frac{R_{
    \rm vdW}^4}{R_C^4}\right)^2. \label{eq:rd_long}
\end{eqnarray}
It is worth pointing out that in the limit of $R \gg R_{\rm vdW}$ the ``long-range'' formula reduces to the ``asymptotic'' one.

To obtain the short range expression we similarly combine the formula for the vibrational spacing [Eq.~(\ref{eq:spacing_rd})], but with the short-range (WKB) wavefunction [Eq.~(\ref{eq:wf_short})], and, in this case, also take into account the WKB phase correction [Eq.~(\ref{eq:spa})]. This has the effect of replacing the $\sin^2[\phi(R_C)]$ term in the model wavefunction with an appropriate $\sin^2[\phi(R_C)+\Delta\beta(R_C,R_t)]$ term:
\begin{eqnarray}
    l_{\rm opt}^{\rm r-d, short} & = & \frac{3\lambda_a^3}{16\pi c} I f_{\rm rot} \frac{2\sqrt{2\mu}}{\hbar \sqrt{\pi}} \frac{\Gamma(4/3)}{\Gamma(5/6)} \frac{1}{\sqrt{-E_b}} \bar a \left[ 1+ (a/{\bar a}-1)^2\right] \frac{R_C^3}{4 R_{\rm vdW}^2} \sin^2\left[\phi(R_C)+\Delta\beta(R_C,R_t)\right]. \label{eq:rd_short}
\end{eqnarray}
\end{subequations}

Similar formulas for a van der Waals dominated excited state can be constructed analogously and the final formulas are:
\begin{subequations}
\begin{eqnarray}
    l_{\rm opt}^{\rm vdW, asym} & = &\frac{3\lambda_a^3}{16\pi c} I f_{\rm rot} \frac{2\sqrt{2\mu}}{\hbar \sqrt{\pi}} \frac{\Gamma(7/6)}{\Gamma(2/3)} \left(\frac{\Delta C_6}{C_6^e}\right)^{1/6} \frac{1}{\sqrt{-E_b}} \left(R_C-a\right)^2, \label{eq:vdw_asym}\\
    l_{\rm opt}^{\rm vdW, long} & = &\frac{3\lambda_a^3}{16\pi c} I f_{\rm rot} \frac{2\sqrt{2\mu}}{\hbar \sqrt{\pi}} \frac{\Gamma(7/6)}{\Gamma(2/3)} \left(\frac{\Delta C_6}{C_6^e}\right)^{1/6} \frac{1}{\sqrt{-E_b}} \left(R_C-a-\frac{8}{15}\frac{R_{\rm vdW}^4}{R_C^3}\right)^2\left(1-\frac{4}{5}\frac{R_{
    \rm vdW}^4}{R_C^4}\right), \label{eq:vdw_long}\\
    l_{\rm opt}^{\rm vdW, short} & = & \frac{3\lambda_a^3}{16\pi c} I f_{\rm rot} \frac{2\sqrt{2\mu}}{\hbar \sqrt{\pi}} \frac{\Gamma(7/6)}{\Gamma(2/3)}
    \left(\frac{\Delta C_6}{C_6^e}\right)^{1/6} \frac{1}{\sqrt{-E_b}} \bar a \left[ 1+ (a/{\bar a}-1)^2\right] \frac{R_C^3}{4 R_{\rm vdW}^2} \sin^2\left[\phi(R_C)+\Delta \beta(R_C, R_t)\right]. \label{eq:vdw_short}
\end{eqnarray}
\end{subequations}
\end{widetext}

\section{Examples}

Now we can proceed to testing our approximations. To test the formulas for the resonant dipole case, we will use the example of intercombination line photoassociation of Yb$_2$~\cite{Tojo2006, Borkowski2009}. For the van der Waals case we will look at photoassociation near the D1 line of Rb in the Rb+Sr system~\cite{Devolder2018, Borkowski2017b}.

\begin{figure}
    \includegraphics[width=0.48\textwidth]{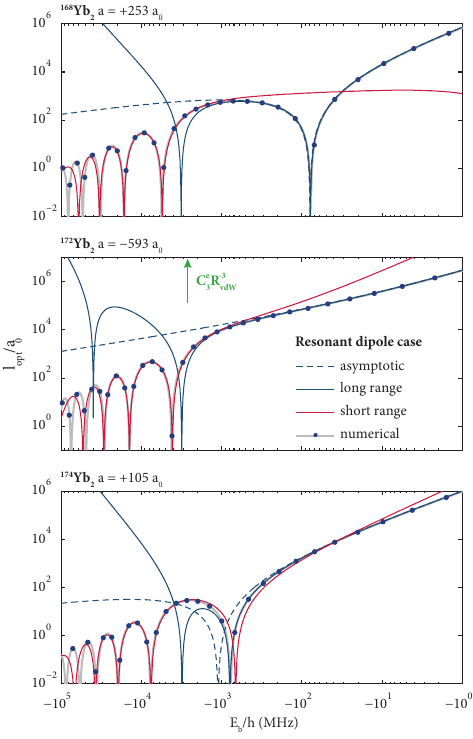}
    \caption{Optical lengths in a resonant-dipole-dominated system on the example of intercombination-line OFRs in Yb~\cite{Tojo2006, Borkowski2009}. Dots represent excited bound states. The grey lines connecting the dots were interpolated by scaling the quantum defect in the excited state potential. The blue dashed and solid lines denote the respective \emph{asymptotic} and \emph{long-range} models, Eqs.~(\ref{eq:rd_asym}) and (\ref{eq:rd_long}). The red lines denote the \emph{short-range} model, Eq.~(\ref{eq:rd_short}). The asymptotic and long-range models match the numerical results down to, respectively, about $-1$~GHz and $-2$~GHz detuning. The enhanced modeling of the ground state wavefunction~\cite{Julienne1996} employed in the long-range model improves the agreement compared to the asymptotic model. For larger detunings the short-range model utilizing the WKB ground state wavefunction correctly reproduces the numerical optical lengths satisfactorily down to about $-20$~GHz beyond which the positions of nodes are no longer correct.
    A more detailed error plot for the $^{174}$Yb$_2$ case (lower panel) is shown in Fig.~\ref{fig:errors}(a).
    \label{fig:lopt_rd}}
\end{figure}

\begin{figure}
    \includegraphics[width=0.48\textwidth]{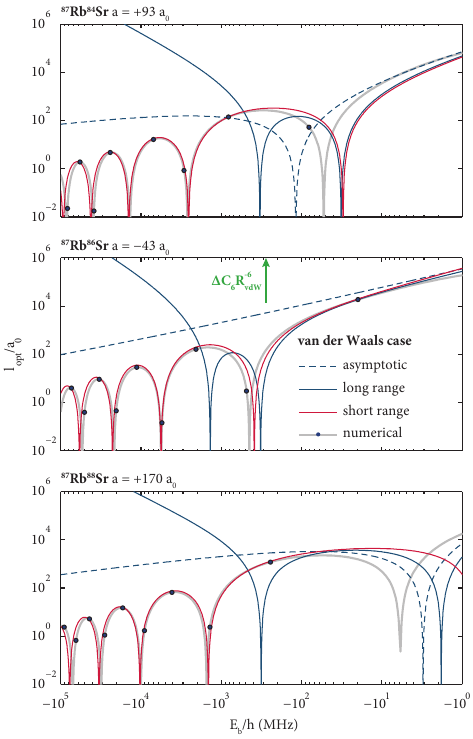}
    \caption{Optical lengths in a van der Waals system on the example of RbSr OFRs near the Rb D1 line. Here all of the numerical values are satisfactorily reproduced by the ``short range'' model, Eq.~(\ref{eq:vdw_short}). The ``asymptotic'' and ``long-range'' models, Eqs.~(\ref{eq:vdw_asym})~and~(\ref{eq:vdw_long}), are applicable only to very weakly bound states. This is easily explained by pointing out that the outer turning points for almost all excited states lie inwards of $R_{\rm vdW}$, where only the short-range WKB ground state wavefunction is applicable. An error plot for the $^{87}$Rb$^{86}$Sr case in the middle panel is shown in Fig.~\ref{fig:errors}(b).
    \label{fig:lopt_vdw}}
\end{figure}

\subsection{A resonant-dipole case: intercombination line photoassociation in Yb}

\begin{figure}[b]
    \includegraphics[width=0.48\textwidth]{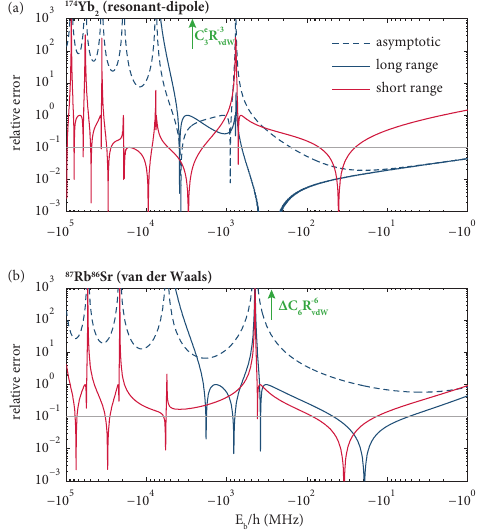}
    \caption{Example relative errors of approximate formulas for the optical lengths of Feshbach resonances: for $^{174}$Yb as a representative example of a resonant-dipole system (a) and a van der Waals system, $^{87}$Rb$^{86}$Sr (b). The blue dashed lines represent the asymptotic models, and blue solid lines are represent the long-range model. The red solid lines denote the short-range models. Finally, the horizontal grey lines mark a 10\% error.}
    \label{fig:errors}
\end{figure}

Our first example is a resonant-dipole dominated case of photoassociation near the $^1$S$_0$$\rightarrow$$^3$P$_1$ transition of Yb~\cite{Tojo2006, Borkowski2009}. The excited $^1$S$_0$$\rightarrow$$^3$P$_1$ asymptote supports four Hund's case~(c) molecular states with the projection of electronic angular momentum on the internuclear axis $|\Omega|$= 0, 1 and with \emph{gerade} (\emph{g}) and \emph{ungerade} (\emph{u}) symmetry. By Laporte rule only electric dipole transitions to the \emph{u} excited states are possible from the \emph{g} electronic ground state. The long-range interaction in the excited electronic state is dominated by the resonant-dipole terms
\begin{equation}
    C_{3, 0}^e = \frac{3}{2}\frac{\hbar}{\tau_A}\left(\frac{\lambda_A}{2\pi}\right)^3 \label{eq:c3}
\end{equation}
for the $|\Omega| = 0$ state and $C_{3,1}^e = -C_{3,0}^e / 2$ for the $|\Omega| = 1$ state. Here $\tau_A$ and $\lambda_A$ are the respective excited state lifetime and transition wavelength. The resonant-dipole interaction is strongly repulsive in the $1_u$ state leaving $0_u^+$ as the only molecular state supporting a series of bound states near the dissociation limit. From an $s$-wave collision only excited states with total angular momentum $J_e=1$ can be reached.

Figure~{\ref{fig:lopt_rd}} shows optical lengths of intercombination line OFRs in the r-d dominated Yb system. The numerical optical lengths were calculated using previous ground~\cite{Borkowski2017a} and excited state models~\cite{Borkowski2009} using the Colbert-Miller discrete variable representation method~\cite{Colbert1992} with a variable grid~\cite{Tiesinga1998}. Both the excited and ground state wavefunctions were calculated using the DVR method. For the scattering wavefunction we picked the lowest solution above the dissociation limit, which we found to be energetically well within the Wigner threshold law. This enabled us to calculate the Franck-Condon factor in Eq.~(\ref{eq:lopt}) as a simple vector dot product between the DVR matrix eigenvectors.

The ground state potential is based on \emph{ab initio} calculations~\cite{Buchachenko2007} with fitted long range parameters, $C_6^g\approx 1937\,E_h a_0^6$, $C_8^g\approx 2.265\times10^5 E_h a_0$ ($E_h$ and $a_0$ are the atomic units of energy and length). The scattering lengths for the $^{168}$Yb, $^{172}$Yb and $^{174}$Yb isotopes are  $+253\,a_0$, $-593\,a_0$ and $+105\,a_0$, respectively~\cite{Kitagawa2008, Borkowski2017a}. The van der Waals length is $R_{\rm vdW} \approx 78\,a_0$ and the mean scattering length $\bar a \approx 74\,a_0$. For the excited state we used a Lennard-Jones+C$_8$ potential model~\cite{Borkowski2009} where the $C_3^e \approx 0.1949\,E_h a_0^3$ is calculated from Eq.~(\ref{eq:c3}) using the lifetime $\tau_A$ = 869.6~ns ($\gamma_a = 2\pi\times183~{\rm kHz}$) and the transition wavelength $\lambda_a=555.8$~nm. The rotational H\"onl-London factor is $f_{\rm rot} = 1/3$~\cite{Machholm2001, Borkowski2009, Nicholson2015} and the excited state has an added rotational interaction energy $V_{\rm rot} = (\hbar^2/2\mu{}R^2)(J_e(J_e+1)+2)$~\cite{Mies1978}.

The tested binding energy range from $-1\,{\rm MHz}$ to $-100\,{\rm GHz}$ corresponds to $R_C$ between over $1000\,a_0$ and $25\,a_0$ and allows us to demonstrate the crossover between the asymptotic, long-range and short-range models and their limitations. The transition between the long-range and short-range models being correct occurs for the range of binding energies where the Condon points lie close to $R_{\rm vdW}$. This is related to the behavior of the ground state scattering wavefunction models and their ranges of validity: $R\gg R_{\rm vdW}$ for the asymptotic, $R\gtrapprox R_{\rm vdW}$ for the long- and $R\lessapprox R_{\rm vdW}$ for the long-range model~\cite{Jones2006}. The ``cutoff'' binding energy that marks the transition from short- to long-range models follows from requiring that the Condon point $R_C \approx R_{\rm vdW}$. This happens when $E_b/h\approx -C_3R_{\rm vdW}/h \approx -2.7\,\rm GHz$ [marked with an arrow in Figures \ref{fig:lopt_rd} and \ref{fig:errors}(a)].

In practice, the asymptotic model works for bound state energies down to about $-1$~GHz. The improved modeling of the scattering wavefunction in the long-range model extends this to about $-2$~GHz. The short range model worked well for energies between about $-2\,{\rm GHz}$ and $-20\,{\rm GHz}$. For bound states below that the contribution of the excited state vdw interaction becomes significant, and our assumption of a pure r-d excited state is no longer valid. If the scattering length $a$ is close to $\bar a$ (like in $^{174}$Yb) both models can slightly misplace the last node because of the influence the vdw potential on the wavefunction around $R_{\rm vdw}$~\cite{Jones2006}. The top panel of figure~\ref{fig:errors} shows the relative errors for the three models for the example case of $^{174}$Yb. The asymptotic model matches the numerical calculations to within $10\%$ for bound state energies down to about $-200$~MHz; the long-range model, again, has an extended region of applicability. Away from nodes the short range model is correct to within about $20\%$ for bound state energies from about $-2$~GHz to about $-20$~GHz. 

\subsection{A van-der-Waals case: Rb D1-line photoassociation of RbSr}

As an example of a system with a van der Waals dominated excited state we pick photoassociation of Rb and Sr atoms near rubidium's 795~nm $^2$S$_{1/2}$$\rightarrow$$^2$P$_{1/2}$ D1 transition~\cite{Devolder2018}. There is only one Hund's case (c) molecular potential near RbSr's $^2$P$_{1/2}$+$^1$S$_0$ asymptote, with total electronic angular momentum $j = 1/2$ and projection $|\Omega| = 1/2$. The case (c) long range interaction coefficient $C_6^e(|\Omega|=1/2)$ can be expressed~\cite{Borkowski2017b} using interaction coefficients for the $^1$S+$^2$P Hund's case (a) $\Sigma$ and $\Pi$ curves: 
\begin{equation}
    C_6^e(|\Omega|=1/2) = \frac{1}{3}C_6^e(\,^2\Sigma) + \frac{2}{3}C_6^e(\,^2\Pi),
\end{equation}
with the respective coefficients for the $^2\Sigma$ and $^2\Pi$ states equal to $23\,324$ and $8\,436$ atomic units. Importantly, the resultant effective $C_6^e(|\Omega|=1/2) = 13\,399$~a.u. is over three times larger than the van der Waals coefficient in the ground state, $C_6^g = 3686\, E_h a_0^6$~\cite{Ciamei2018}. Having the van der Waals coefficient of the excited state much larger than that in the ground state is a prerequisite for using the stationary phase approximation, Eq.~(\ref{eq:spa}). The breakdown of this approximation when the two interaction coefficients are of similar magnitude is described in detail in Ref.~\cite{Ciurylo2006}.

To calculate the optical lengths in this system numerically we used the recent empirical Lennard-Jones+C$_8$ ground state potential~\cite{Ciamei2018} with $C_6^g \approx 3686\, E_h a_0^6$ and $C_8^g \approx 4.64\times10^5 E_h a_0^8$, whereas for the $j=1/2,\,\Omega=1/2$ excited state we used a Lennard-Jones potential whose depth matches that of the \emph{ab initio} (2)~$\Omega=1/2$ potential in~\cite{Devolder2018}. We assumed the ground state collision to be dominated by the $s$-wave, with no rotational factor in the interaction potential, and we assumed transitions to the rotationless $J=1/2$ excited state~\cite{Borkowski2017b}. The parameters required to use the approximate formulas, Eqs.~(\ref{eq:vdw_asym})--(\ref{eq:vdw_short}) are as follows. The scattering lengths for $^{87}$Rb paired with $^{84}$Sr, $^{86}$Sr, and $^{88}$Sr are $+93\,a_0$, $-43\,a_0$ and $+170\,a_0$~\cite{Barbe2018, Ciamei2018}, $R_{\rm vdW} \approx 77.5\,a_0$, $\bar a \approx 74\,a_0$ and finally the rotational H\"onl-London factor is $f_{\rm rot}=1$~\cite{Borkowski2017b}. 

Figure~\ref{fig:lopt_vdw} shows the numerically calculated optical lengths for three isotopic combinations of the RbSr system and their approximate counterparts. Unlike the resonant-dipole example in the previous section, here the short range model describes virtually all resonances from the dissociation limit down to $E_b/h\approx -50\,{\rm GHz}$. The utility of the long range and asymptotic models is limited, as for transitions to most bound states the Condon point $R_C$ lies at much shorter internuclear separations than the van der Waals radius $R_{\rm vdW}$. In fact, for RbSr a Condon point at $R_{\rm vdW}\approx 77.5\,a_0$ corresponds to a excited state binding energy of only about $-\Delta C_6 R_{\rm vdW}^{-6}/h = -290\,{\rm MHz}$ [Figs.~\ref{fig:lopt_vdw} and \ref{fig:errors}(b)]. Van der Waals systems usually have at most one or two bound states this close to the dissociation limit, so we expect the long range model to occasionally be applicable to the most weakly bound state in a van der Waals system. Figure~\ref{fig:errors}(b) shows the relative errors of each of the approximate formulas. Compared to the resonant-dipole example, here the formulas are more qualitative. While the short range formula still reproduces the numerical results to within about 30\% (away from nodes), the long-range model fails below about $-300$~MHz and the asymptotic model fails everywhere.  

\section{The optical length in context of associative STIRAP}

\begin{figure}
    \includegraphics[width=0.48\textwidth]{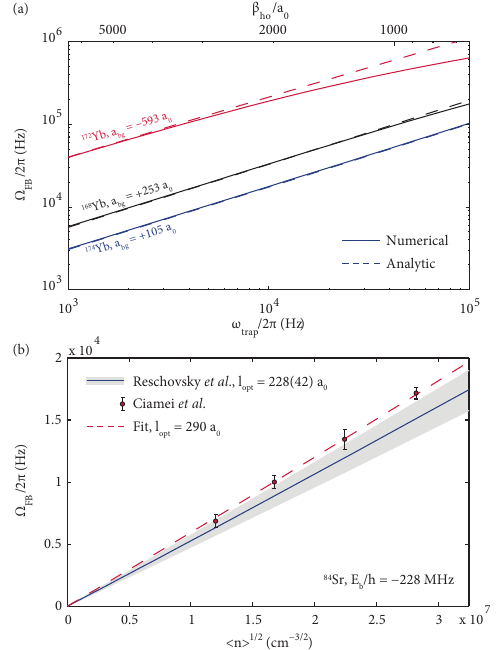}
    \caption{Calculation of free-bound Rabi frequencies $\Omega_{\rm FB}$ from the optical length (for $I=1\,{\rm W/cm^2}$). (a) Numerical and analytic [Eq.~(\ref{eq:omega_FB})] Rabi frequencies for transitions to $^1$S$_0$+$^3$P$_1$ states at $-310$~MHz, $-353$~MHz and $-303$~MHz in $^{168}$Yb, $^{172}$Yb, and $^{174}$Yb, respectively, as a function of trapping frequency $\omega_{\rm trap}$. Equation~(\ref{eq:omega_FB}) remains accurate as long as $a \ll \beta_{\rm ho}$ (alternative axis). (b) $\Omega_{\rm FB}$ for the $-228$-MHz $^1$S$_0$+$^3$P$_1$ state in $^{84}$Sr as a function of $\langle n \rangle^{1/2}$. Data points were measured by Ciamei~\emph{et al.}~\cite{Ciamei2017}, the shaded area is calculated from $l_{\rm opt} = 228(42)\,a_0$ measured by Reschovsky~\emph{et al.}~\cite{Reschovsky2018}.}
    \label{fig:omegafb}
\end{figure}

\subsection{Relationship between the optical length and free-bound transitions in a 3D optical lattice site}

The utility of $l_{\rm opt}$ can be extended to coherent molecule production via associative STIRAP~\cite{Bergmann1998, Ni2008, Vitanov2017} in a doubly occupied Mott insulator~\cite{Stellmer2012, Ciamei2017}. Here we will give an expression for the ``free-bound'' Rabi frequency $\Omega_{\rm FB}$, induced when a laser couples an initially unbound atomic pair in a 3D optical lattice site to an excited molecular state, in terms of $l_{\rm opt}$. This is useful as $\Omega_{\rm FB}$ depends on both the molecular physics as well as the trap parameters, while $l_{\rm opt}$ is an intrinsically molecular quantity. For an atomic pair with similar masses and trapping frequencies $\omega_{\rm trap}$ the centre-of-mass and relative motion separate~\cite{Busch1998, Chen2007}. The latter is governed by a radial Schr\"odinger equation for the previous potential $V_g(R)$, but with an added harmonic potential $V_{\rm ho}(R)= \frac{1}{2}\mu \omega_{\rm trap}^2 R^2$. The weak trapping potential quantizes the scattering continuum into discrete trap states separated by $\sim 2 \hbar \omega_{\rm trap}$ and whose positions are the solutions of~\cite{Busch1998, Block2002, Shresta2005a, Chen2007}:
\begin{equation}
    \frac{1}{2}\frac{\Gamma(1/4-e/2)}{\Gamma(3/4-e/2)} = \frac{a}{\beta_{\rm ho}}, \label{eq:busch}
\end{equation}
where $e=E_{\rm trap}/\hbar\omega_{\rm trap}$ and $\beta_{\rm ho} = \sqrt{\hbar/\mu\omega_{\rm trap}}$ is a characteristic length associated with the harmonic trap potential, typically on the order of $10^3-10^4\,a_0$.

In analogy to the OFR stimulated width $\hbar \Gamma_{\rm stim} = 2\pi\rvert\langle f_g \rvert V^{\rm opt} \rvert \psi_b \rangle\rvert^2$, the ``free-bound'' Rabi frequency may be defined as $\hbar \Omega_{\rm FB} =\, \rvert \langle \psi_{\rm trap} \rvert V^{\rm opt} \rvert \psi_b \rangle \rvert$, where $\psi_{\rm trap}$ is the trap state wavefunction and $V^{\rm opt}$ is the optical coupling matrix element~\cite{Bohn1999, Nicholson2015}. At internuclear distances that contribute to the Franck-Condon factor -- typically much shorter than $\beta_{\rm ho}$ -- the trapping potential is weak compared to the trap state energy. Since $\psi_{\rm trap}$ and $f_g$ are the solutions of radial Schr\"odinger equations that differ only by the weak harmonic potential that vanishes for small $R$, the ``trap'' wavefunction can be approximated to within a scaling factor by the scattering wavefunction calculated for the trap state energy. The scaling factor can be taken from MQDT~\cite{Mies1984, Bohn1999}: $\psi_{\rm trap} = (\partial E_{\rm trap}/\partial \nu)^{1/2} f_g(k_{\rm trap})$, where $(\partial E_{\rm trap}/\partial \nu)$ is the trap state spacing. Finally, we recall the relationship between the stimulated width and the optical length, $\Gamma_{\rm stim} = 2 k_{\rm trap} l_{\rm opt} \gamma_m$, for the wavenumber $k_{\rm trap}=\sqrt{2\mu E_{\rm trap}}/\hbar$. As a result, 
\begin{equation}
    \Omega_{\rm FB} =  \left[\frac{1}{2\pi\hbar} \frac{\partial E_{\rm trap}}{\partial \nu} 2 k_{\rm trap} l_{\rm opt} \gamma_{m}\right]^{1/2}. \label{eq:omega_FB}
\end{equation}
In a Mott insulator~\cite{Ciamei2017} the atoms occupy the lowest trap state above the dissociation limit, so we use $k_{\rm trap} = \sqrt{2 e_0}/\beta_{\rm ho}$ and $(\partial E_{\rm trap}/\partial \nu) \approx \hbar \omega_{\rm trap} (e_1-e_0)$, where $e_0$ and $e_1$ are the two lowest solutions of Eq.~(\ref{eq:busch}). We stress that this derivation does not need any of the assumptions we previously made for our approximate formulas for $l_{\rm opt}$, but only that $\beta_{\rm ho}$ is much larger than any other length scale, particularly $a$.

\subsection{Results}

Numerical testing for Yb shows that for typical trap frequencies Eq.~(\ref{eq:omega_FB}) works with an accuracy better than 10\% unless $a$ is appreciable compared to $\beta_{\rm ho}$ [Fig.~\ref{fig:omegafb}(a)]. In $^{174}$Yb characterized by a moderate scattering length of $a = 105~a_0$, the agreement is to better than 2\% for all tested $\omega_{\rm trap}$. In fact, as long as $\rvert a/\beta_{\rm ho}\rvert \leq 0.1$, this accuracy is retained for all tested isotopes. If the scattering length is resonant, as in $^{172}$Yb ($a=-593\,a_0$), our model becomes less accurate: for a trapping frequency of $2\pi\times10$~kHz ($\rvert a/\beta_{\rm ho} \rvert\approx 0.3$) the accuracy deteriorates to about 10\%.

Recent experimental investigations of associative STIRAP in $^{84}$Sr Mott insulator by the Amsterdam group~\cite{Ciamei2017} and, independently, of photoassociation rates in a $^{84}$Sr BEC in a dipole trap at JQI~\cite{Reschovsky2018} allow for a real-world test of Eq.~(\ref{eq:omega_FB}). The Amsterdam group measured $\Omega_{\rm FB}$ for transitions to the $-228$~MHz state near the $^1$S$_0$+$^3$P$_1$ asymptote in $^{84}$Sr$_2$ [Fig.~\ref{fig:omegafb}(b)] and found the Rabi frequency to be proportional to the square root of the single-atom average on-site density $\langle n \rangle=1/(2\pi)^{3/2}a_{\rm ho}^3$, where $a_{\rm ho} = \sqrt{\hbar/m \omega_{\rm trap}}=\beta_{\rm ho}/\sqrt{2}$. Indeed, aside from the weak dependence of reduced trap energies on the scattering length, the trap state spacing $\partial E/\partial n \propto \omega_{\rm trap}$, the wavenumber $k_{\rm trap}\propto \omega_{\rm trap}^{1/2}$ and therefore the free-bound Rabi frequency $\Omega_{\rm FB} \propto l_{\rm opt}^{1/2} \omega_{\rm trap}^{3/4} \propto l_{\rm opt}^{1/2} \langle n\rangle^{1/2}$. By fitting Eq.~(\ref{eq:omega_FB}) to the measured $\Omega_{\rm FB}$ we extract the optical length $l_{\rm opt} = 290(13)\,a_0$. Here $a=122.7\,a_0$~\cite{MartinezDeEscobar2008} and $\gamma_m = 2\times2\pi\times7.5$~kHz~\cite{Skomorowski2012, Borkowski2014}; parentheses indicate the statistical fit uncertainty. The extracted optical length agrees to within 1.4 mutual sigma with the experimental $l_{\rm opt} = 228(42)\,a_0$ measured by the JQI group~\cite{Reschovsky2018}. Additionally, theoretical $\Omega_{\rm FB}$ calculated using the JQI optical length [shaded area in Fig.~\ref{fig:omegafb}(b)] generally reproduce the measured $\Omega_{\rm FB}$.

\section{Conclusion}
In conclusion, we have developed simple analytic formulas for the optical Feshbach resonance strength parameter, the optical length, for near-threshold bound states using the stationary phase approximation~\cite{Jablonski1945, Julienne1996, Bohn1999, Boisseau2000, Ciurylo2006}. We rely on the excited state potential being dominated by either a resonant-dipole $R^{-3}$ interaction typical for homonuclear photoassociation near strong lines, or a van der Waals $R^{-6}$ tail appropriate for heteronuclear systems. The optical length is expressed in terms of dominant interaction parameters and the $s$-wave scattering length. We have demonstrated our model using Yb$_2$ and RbSr as real-world examples and found semi-quantitative agreement for resonances up to tens of GHz from the dissociation limit. The derived expressions could aid the design of future photoassociation or OFR experiments when only the long range interaction parameters are known. The formulas could also potentially be used to determine the scattering length from experimental spectra if a more accurate method like two-color photoassociation spectroscopy is not available or practical. The resonant-dipole formulas will work for homonuclear OFRs near any allowed atomic transition, but have worked well for intercombination line OFRs in Yb~\cite{Tojo2006, Borkowski2009, Kim2016} and should apply to thus far unexplored systems with similarly strong intercombination lines, particularly Hg~\cite{Krosnicki2015, Yamanaka2015, Tyumenev2016}, and Cd~\cite{Maslowski2009, Urbanczyk2017, Yamaguchi2019, Dzuba2019} considered as references in optical lattice clocks. 

We have also shown how the optical $l_{\rm opt}$ may be used in the context of coherent molecular formation via associative STIRAP in a 3D optical lattice~\cite{Vitanov2017, Ciamei2017} as a measure of transition strength that is independent of trap parameters. We have found the ``pump'' beam Rabi frequency $\Omega_{\rm FB}$ to be proportional to ${l_{\rm opt}^{1/2}}$ and approximately proportional to on-site density $\left< n \right>^{1/2}$ corroborating the empirical observation of Ciamei~\emph{et al.}~\cite{Ciamei2017} for the $^1$S$_0$+$^3$P$_1$ $0_u^+$ $-228$-MHz resonance near the intercombination line in $^{84}$Sr. From their experimental $\Omega_{\rm FB}$ we extracted a value of $l_{\rm opt} = 290(13)\,a_0$, which agrees with an independently measured $l_{\rm opt} = 228(42)\,a_0$ of Reschovsky~\emph{et al.}~\cite{Reschovsky2018}. 

\begin{acknowledgments}
I would like to thank Piotr Żuchowski, Roman Ciuryło, Paul S. Julienne and Alessio Ciamei for useful discussions. I acknowledge support from the National Science Centre (Grant No. 2017/25/B/ST4/01486).  This work is part of an ongoing research program of the National Laboratory FAMO in Toruń, Poland.  Calculations have been carried out at the Wroclaw Centre for Networking and Supercomputing (http://www.wcss.pl), Grant No. 353.
\end{acknowledgments}
\bibliography{library}
\end{document}